\documentclass[conference]{IEEEtran}
\IEEEoverridecommandlockouts
\usepackage{cite}
\usepackage{amsmath,amssymb,amsfonts}
\usepackage{algorithmic}
\usepackage{graphicx}
\usepackage{textcomp}
\usepackage{xcolor}

\usepackage[normalem]{ulem}
\usepackage{float}
\usepackage{multirow}
\usepackage{physics}
\usepackage{listings}
\usepackage{booktabs}
\usepackage{xspace}
\usepackage{multicol}
\usepackage{tabularx}
\usepackage{color, colortbl}
\usepackage{balance}
\usepackage{url}
\def\BibTeX{{\rm B\kern-.05em{\sc i\kern-.025em b}\kern-.08em
    T\kern-.1667em\lower.7ex\hbox{E}\kern-.125emX}}
    
\newcommand{\projname} {{MosaicSim}}
\newcommand{\myfig} {{Figure}}

\definecolor{Gray}{gray}{0.9}
\newcolumntype{L}[1]{>{\raggedright\let\newline\\\arraybackslash\hspace{0pt}}m{#1}}
\newcolumntype{R}[1]{>{\raggedleft\let\newline\\\arraybackslash\hspace{0pt}}m{#1}}
\newcolumntype{Z}{>{\raggedleft\let\newline\\\arraybackslash\hspace{0pt}}X}

\colorlet{shadecolor}{gray!40}

\newcommand*{\aboverulesepcolor}[1]{%
  \noalign{%
    \begingroup
      \color{#1}%
      \hrule height\aboverulesep
    \endgroup
    \kern-\aboverulesep
  }                               
}

\begin{document}

\title{\projname: A Lightweight, Modular Simulator for Heterogeneous Systems\\
\thanks{This version is a variation of the original paper published in \textit{The 2020 IEEE International Symposium on Performance Analysis of Systems and Software} that includes additional results demonstrating the accuracy of \projname's memory hierarchy performance modeling.}
}

\author{\IEEEauthorblockN{Opeoluwa Matthews\IEEEauthorrefmark{1} \hspace{0.5cm} Aninda Manocha\IEEEauthorrefmark{1} \hspace{0.5cm} Davide Giri\IEEEauthorrefmark{2} \hspace{0.5cm} Marcelo Orenes-Vera\IEEEauthorrefmark{1} \hspace{0.5cm} Esin Tureci\IEEEauthorrefmark{1} \\ Tyler Sorensen\IEEEauthorrefmark{1}\IEEEauthorrefmark{3} \hspace{0.5cm} Tae Jun Ham\IEEEauthorrefmark{4} \hspace{0.5cm} Juan L. Aragón\IEEEauthorrefmark{5} \hspace{0.5cm} Luca P. Carloni\IEEEauthorrefmark{2} \hspace{0.5cm} Margaret Martonosi\IEEEauthorrefmark{1}} \\ \normalsize \IEEEauthorrefmark{1}Princeton University \hspace{0.4cm} \IEEEauthorrefmark{2}Columbia University \hspace{0.4cm}\IEEEauthorrefmark{3}UC Santa Cruz \hspace{0.4cm} \IEEEauthorrefmark{4}Seoul National University \hspace{0.4cm} \IEEEauthorrefmark{5}University of Murcia}

\maketitle

\begin{abstract}
As Moore's Law has slowed and Dennard Scaling has ended, architects are increasingly turning to heterogeneous parallelism and domain-specific hardware-software co-designs. These trends present new challenges for simulation-based performance assessments that are central to early-stage architectural exploration. Simulators must be lightweight to support rich heterogeneous combinations of general purpose cores and specialized processing units. They must also support agile exploration of hardware-software co-design, i.e. changes in the programming model, compiler, ISA, and specialized hardware. 

To meet these challenges, we introduce \projname{}, a lightweight, modular simulator for heterogeneous systems, offering accuracy and agility designed specifically for hardware-software co-design explorations.
By integrating the LLVM toolchain, \projname{} enables efficient modeling of instruction dependencies and flexible additions across the stack. Its modularity also allows the composition and integration of different hardware components.
We first demonstrate that \projname{} captures architectural bottlenecks in applications, and accurately models both scaling trends in a multicore setting and accelerator behavior. We then present two case-studies where \projname{} enables straightforward design space explorations for emerging systems, i.e. data science application acceleration and heterogeneous parallel architectures.
\end{abstract}




\section{Introduction}

\begin{figure}[t]
\centering  
\includegraphics[width=\linewidth]{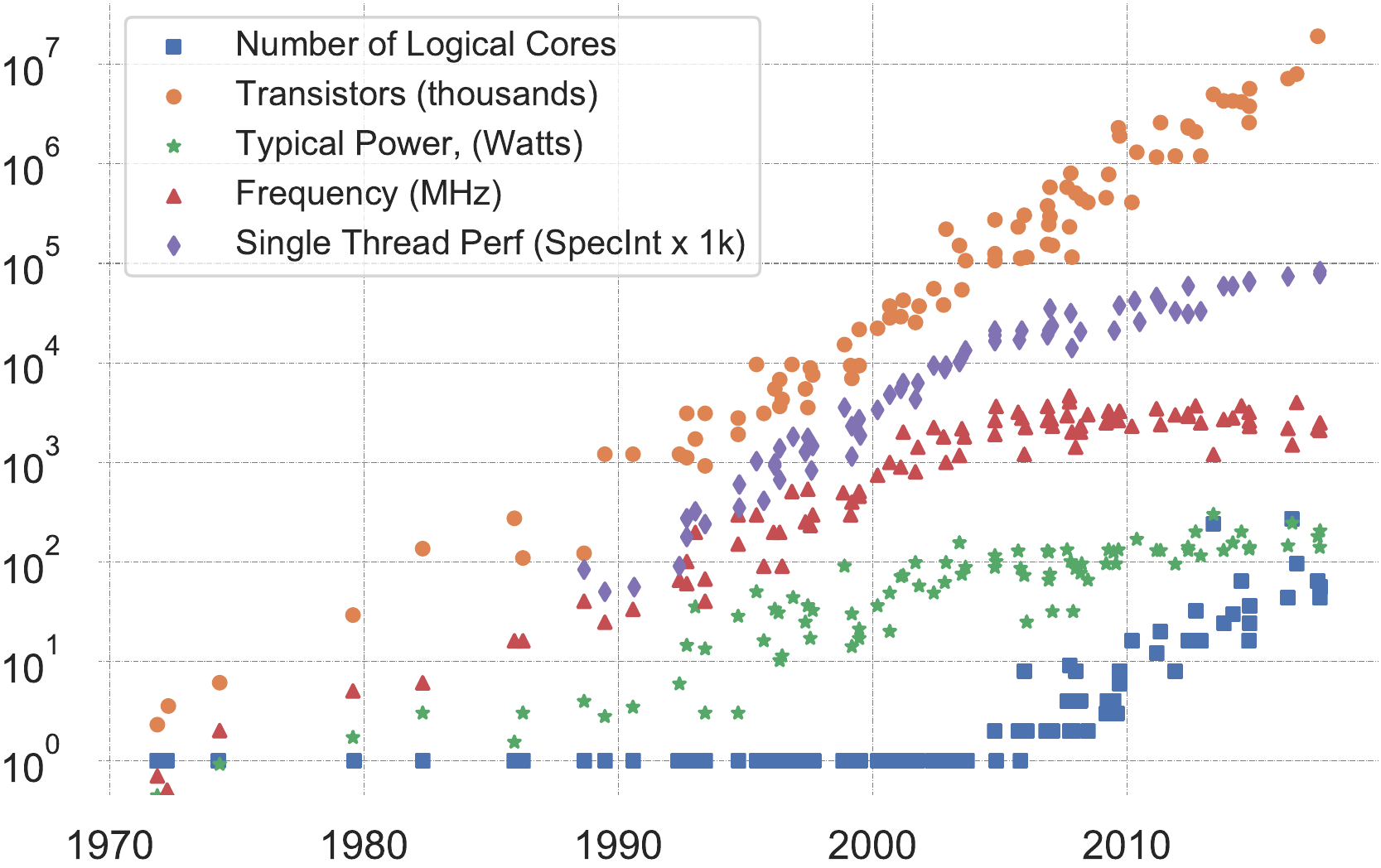}
\caption{42 years of trend data for microprocessor characteristics, graph recreated using data from~\cite{40y} }
\label{fig:42y}
\end{figure}

The last decade has seen a trend of increasing parallelism as a response to the ending of Moore's Law and Dennard scaling. \myfig~\ref{fig:42y} presents microprocessor trends over the past few decades. As computing frequency (red triangles) has plateaued, the number of logical cores (blue squares) has increased. The stagnation in raw computing frequency has also triggered the usage of specialized systems, including heterogeneous architectures and hardware-software co-design, to meet the demands of today's aggressive performance and power goals.
Designers of modern systems are therefore employing combinations of distinct computation elements, including small, low-power cores and high-performing hardware accelerators~\cite{piton, aladdin, Venkat:ISA}.

In their Turing award lecture, Hennessy and Patterson describe a New Golden Age for Computer Architecture, where future performance improvement opportunities encourage vertically-integrated system designs~\cite{goldenage}. Such systems require innovation in programming models, compilers, specialized hardware, and ISAs. Hence, the system design space has seen a Cambrian explosion in diversity at all levels of the stack, necessitating flexible tools for design space exploration.

Well-known simulators, e.g.\ gem5 \cite{gem5}, offer detailed simulation infrastructures for conventional microarchitectures, but make it difficult for a designer to explore changes across other layers of the stack, which have increasing influence in the performance of systems today. 
Other approaches resort to high-level simulation (e.g.\ 1-IPC models or interval simulation~\cite{interval-sim}) that do not accurately capture critical memory bottlenecks of many modern data-intensive applications. 

We present \projname{}, a simulation approach that allows for the exploration of optimizations across the hardware-software stack, while providing accurate modeling of performance bottlenecks and application characteristics. To achieve these goals, we leverage the LLVM framework~\cite{llvm}, which allows us to utilize a mature compiler infrastructure to capture instruction dependencies and collect memory traces. \projname{} executes LLVM IR, which enables ISA-agnostic simulation and supports flexible programming models through compiler passes and specialized instructions. \projname{} then simulates the LLVM IR instructions on modular tile models, which enables straightforward design space exploration of heterogeneous systems. Furthermore, tile modules support a flexible communication model, which allows data-supply hardware-software co-designs to be evaluated. 

To summarize, \projname{} is a lightweight, modular simulator for heterogeneous and hardware-software co-design systems. Its main contributions are:
\begin{itemize}
    \item Enabling flexible programming models, compiler techniques and ISAs through integration with LLVM.
    \item Abstract tile models, which capture key performance microarchitectural details for a variety of core models and accelerators. 
    \item A holistic simulation approach that allows for the composition of different hardware tiles and communication structures, allowing for the simulation of complex heterogeneous systems. 
\end{itemize}

\noindent
We evaluate \projname{} and demonstrate that it:

\begin{itemize}
    \item Accurately captures trends on existing parallel architectures and accelerators (Section~\ref{sec:evaluation}). 
    \item Is able to simulate complex heterogeneous systems, illustrated through three case studies (Section~\ref{sec:case-studies}).
\end{itemize}





\section{\projname{} Overview}
This section describes an overview of the \projname{} simulation methodology. At a high level, \projname{} provides tile-based models of different hardware units, including cores, accelerators, and caches, with an \textit{Interleaver} that composes their behaviors to provide total system estimates.  

\begin{figure}[!t]
\centering  
\includegraphics[width=\linewidth]{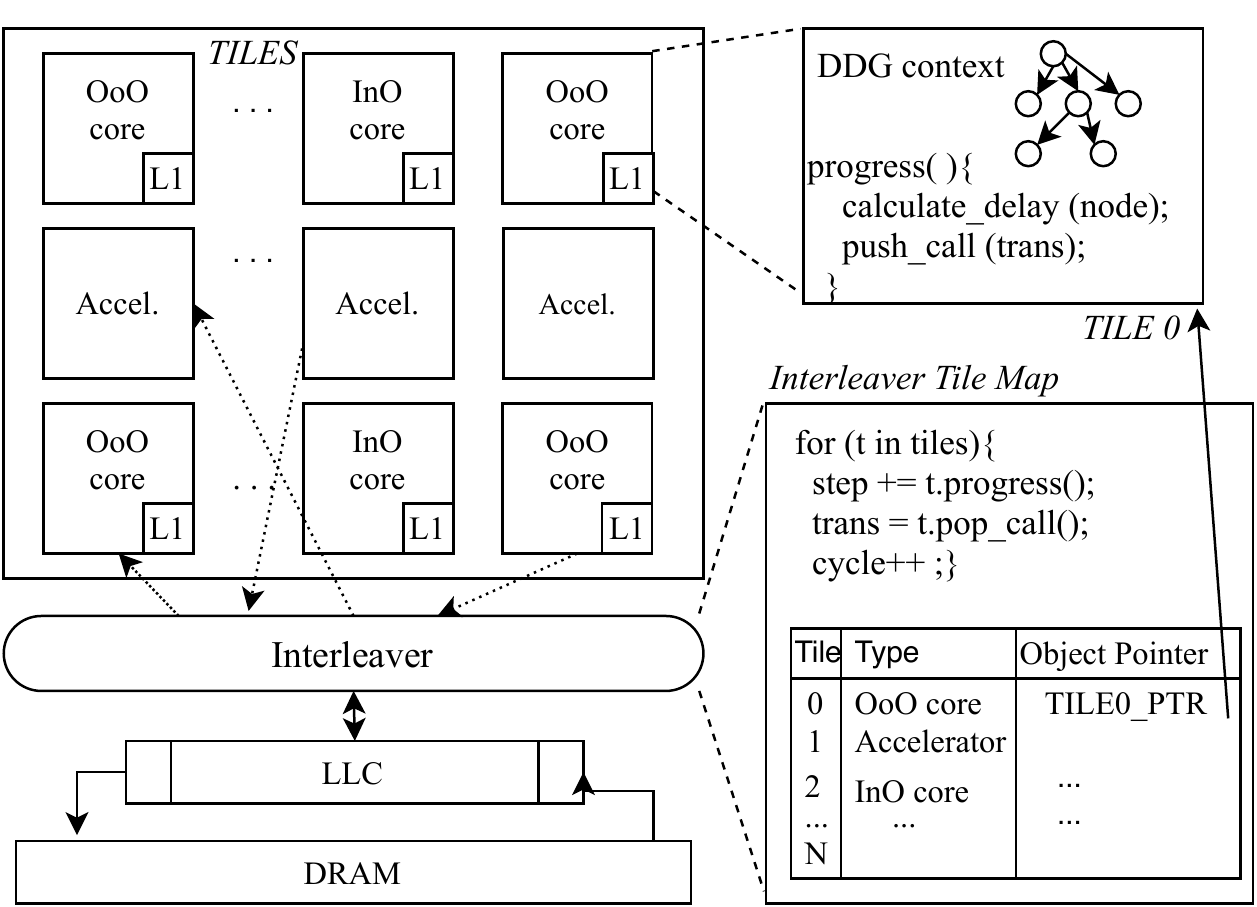}
\caption{\projname{} integrates different modules (e.g.\ CPUs and accelerators) via the Interleaver, which combines module behaviors into system-wide performance estimates.}
\label{fig:digester}
\end{figure}

{\bf Tiled System Model:} Figure~\ref{fig:digester} displays the  tiled system model in \projname{}. The overall design represents an SoC comprised of CPU and accelerator tiles. Each tile in the SoC has a model of its events that contribute to the performance and power of the entire system, and the Interleaver coordinates the interactions of events from different tiles. \projname{} simulates a simple homogeneous chip multiprocessor by instantiating several CPU tiles and allowing the Interleaver to coordinate their interactions, e.g.\ coordinating shared memory hierarchy behavior. Additionally, \projname{} can simulate more heterogeneous processors by providing (and hence, interleaving) more diverse models.  As a design process evolves, accelerators or other specialized hardware can be incorporated as Section~\ref{sec:accel} describes. The direct linkage with an LLVM-based compiler allows straightforward decomposition into models for different units.

As an early-stage tool that targets design space exploration for hardware-software co-design, \projname{} focuses on kernel simulation. This allows for modeling compute or memory bottlenecks in order to provide hardware designers with the necessary insight to make design decisions (e.g. employing accelerators) accordingly. \projname{} is not restricted to kernel modeling and can simulate arbitrary codes as long as LLVM-IR can be obtained. However, full application simulation requires performance models that are often only available in later design stages, e.g.\ filesystem I/O and system calls. 

{\bf Timing Integration:} Distinct tiles may use different notions of execution timing and are modeled to operate concurrently. The Interleaver queries tiles to advance them through the next time unit of execution. Tiles may run at different clock speeds, so the Interleaver queries and coordinates their events accordingly. To communicate, tiles create inter-tile events and enqueue them for the Interleaver to manage. The Interleaver is then responsible for sending a transaction to its destination tile at the right time; it does so by explicitly invoking a destination tile to receive and process message events. 

{\bf Compiler and Software-Hardware Interface:}
\projname{} uses LLVM IR as its ISA, so it is closely integrated with a mature and open-source compiler framework. Wide support for LLVM frontends allows the compiler to take inputs from a variety of languages. While \projname{}'s most developed front-end is C/C++ through Clang~\cite{CLANG}, we also have prototype support for Python (via Numba~\cite{numba}) and performance modeling for TensorFlow Keras~\cite{keras}. The compiler allows further programmer directives to guide hardware components to simulate. For example, the programmer can utilize an accelerator API with common functions (e.g.\ matrix multiplication) to invoke an accelerator model for specific compute tasks, thus allowing the exploration of design performance trade-offs. New instructions, programming paradigms, and pragmas can be straightforwardly added as functions calls identified through LLVM passes. Relevant parameters can then be relayed to the simulator through traces. 
The compiler generates dependency graphs of LLVM IR that the simulator can map onto distinct tiles or analyze for lightweight performance estimation. 

\subsection{Lightweight Tile Models using Dependence Graphs}
\label{sec:graphstomodels}

Tile models begin as abstract models based on data dependence graphs derived from LLVM IR. Namely, from a full software application written in a compatible language, the compiler can identify kernels for which to perform dependence analysis to create a graph-based model. Section~\ref{sec:cpu} describes how such models can account for different hardware characteristics to reflect issue width, in- or out-of-order execution, and other processor attributes.

{\bf Execution Modeling:}
In graph-based tile models, a {\em node} corresponds to a static operation (instruction) and keeps track of its dependents and parents\footnote{We use {\em node} and {\em instructions} interchangeably.}. Dependence analysis is performed to identify basic blocks, which are single-entry, single-exit collections of static instructions in LLVM~\cite{LLVM-basic-block}. Each basic block can have many dynamic instances (e.g.\ when a basic block is executed repeatedly in a \texttt{for} loop), so we call such instances {\em Dynamic Basic Blocks} (DBBs). In a \texttt{for} loop, the basic block remains the same each iteration, but the iteration variable maps to different values, creating different DBBs. {\em Terminator nodes}, or exit points (e.g.\ jump instructions), have edges to DBBs that could be executed next.

\begin{figure*}[!t]
\centering  
\includegraphics[width=0.85\textwidth]{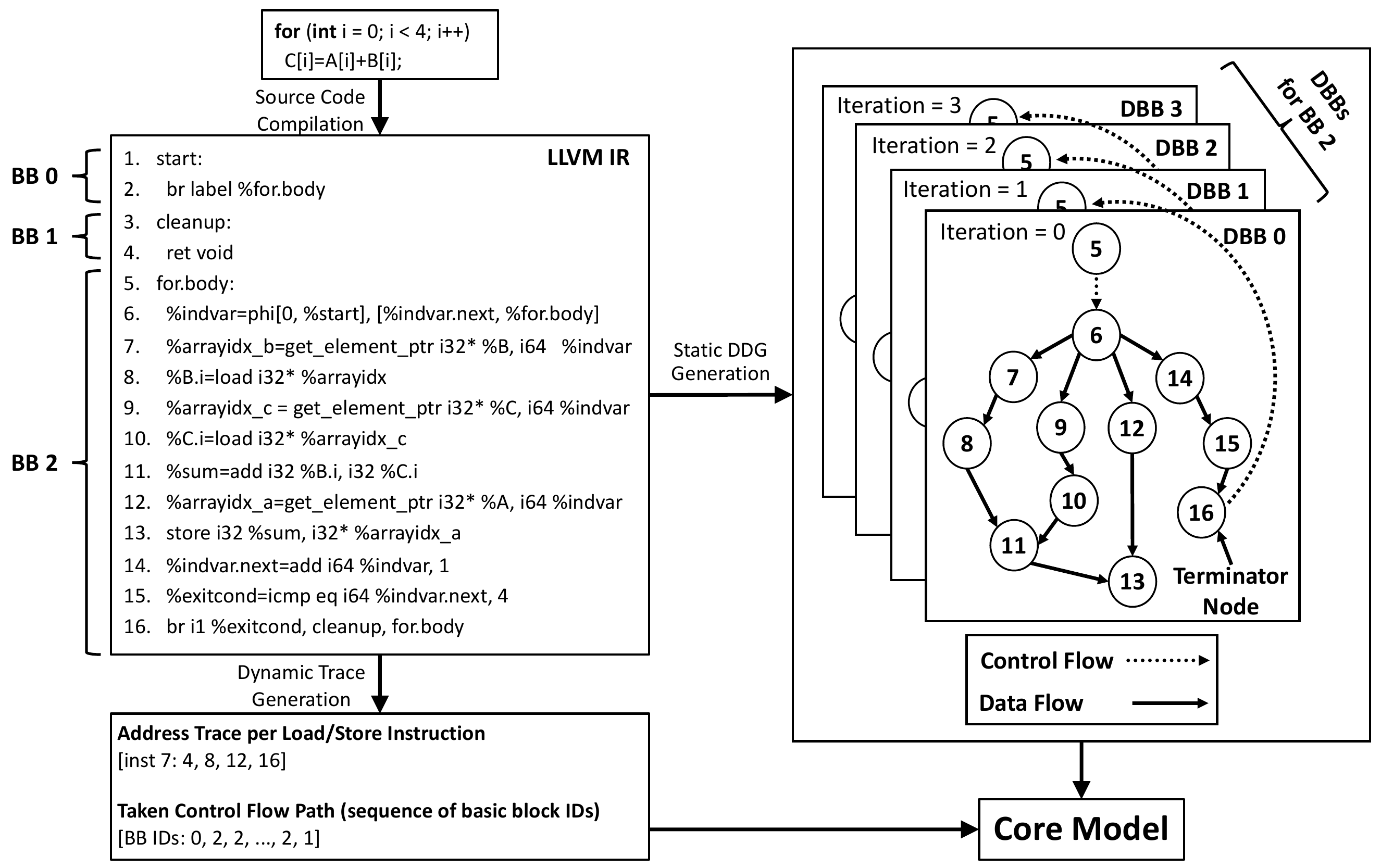}
\caption{MosaicSim's Execution Modeling Flow}
\label{fig:tool_flow}
\end{figure*}

In order to perform dependence analysis, MosaicSim relies on (1)~a Static \textit{Data Dependency Graph} (DDG) Generator and (2)~a \textit{Dynamic Trace Generator} (DTG). Both tools operate on compiled source code with the kernel annotated. The static DDG Generator uses a series of LLVM passes to capture static inter-instruction dependencies and provide a graphical representation of the source code. This representation can involve many DBBs. 
Figure~\ref{fig:tool_flow} illustrates \projname{}'s execution modeling for a core to run a non-speculative example.
Nodes in DBBs correspond to instructions, while edges capture data and control flows within and across DBBs.

Since memory dependencies and control flow paths cannot be completely determined statically, the DTG uses an LLVM pass to create an instrumented x86 executable that, when run, writes two trace files: (1)~a control flow trace that records the dynamic control flow decisions; and (2)~a memory trace that records the addresses for each memory access. Following the native run on the host machine, MosaicSim uses these trace files in the core model, allowing cycle-driven simulation of different execution possibilities (e.g.\ in-order vs. out-of-order).


{\bf Data Dependencies:} The DTG outputs information on all addresses accessed, but address aliasing can occur until the program actually resolves the addresses. Thus, \projname{} implements a \textit{Memory Address Orderer} (MAO), to ensure that true memory dependencies (i.e.\ Read-After-Write dependencies) are respected. The MAO is populated with memory operations in program order, and can be instantiated with various parameters, e.g. to model a traditional Load-Store Queue (LSQ) in core models (see Section~\ref{sec:cpu}).

Before a store instruction executes, it checks the MAO to ensure that there exists no incomplete older memory access with a matching or unresolved address. A load only needs to ensure that there exists no incomplete older store with a matching or unresolved address. If these conditions are not met, the memory operation and its dependent instructions stall. 


{\bf Control Flow Dependencies:}
\projname{} serially launches DBBs based on the control flow path trace and the amount of resources devoted to the core model (see Section~\ref{sec:resource_limits}). Since multiple tiles each run multiple DBBs, the Interleaver coordinates event timing and communication among tiles (detailed in Section~\ref{sec:inter-tile-comm}) and with the memory hierarchy.  

The DTG provides a list of basic block IDs in execution order. For each ID in the list, \projname{} 
launches a new DBB based on the corresponding static basic block. A DBB becomes \textit{live}, or is launched, only after the terminator node that branches to it has completed. Instructions cannot execute unless the DBB they belong to has been launched. Note, however, that despite the serial launching of DBBs, \projname{} can have multiple live DBBs at a given time because a terminator node is not necessarily the last instruction to be completed. For example, terminator node \textcircled{\scriptsize{\textbf{16}}} in Figure~\ref{fig:tool_flow} can be reached in just 5 cycles, but it may take longer to reach node \textcircled{\scriptsize{\textbf{12}}}. Thus, new DBBs for a particular basic block are launched when the terminator node has been reached regardless of whether the current DBB has finished. This leads to a variable number of in-flight DBBs per static basic block. 


In summary, \projname{} enforces the following rules to respect data and control flow dependencies: 

\begin{enumerate}
    \item An instruction cannot be issued unless its DBB has been created {\em and} all of its parent nodes have completed. 
    \item When an instruction completes, MosaicSim attempts to issue its dependents, while also decreasing the dependents' count of uncompleted parents. Dependents with no additional uncompleted parents can be issued (subject to hardware resource constraints, as discussed in Section~\ref{sec:resource_limits}).
    \item When a terminator node completes and if resource limits have not been reached, the Interleaver launches the next DBB based on the control flow path trace from the DTG.
\end{enumerate}


\subsection{Task to Tile Mappings \label{sec:tasktotile}}
A tile executes a \emph{kernel}, which is given as a specially named LLVM function. Different kernels can be mapped to different tiles if distinct DDGs and traces are generated. 


Currently, \projname{} provides a single program, multiple data (SPMD) approach. That is, the user writes one kernel function $K$ and queries a unique tile ID and number of tiles from the execution environment. This provides a familiar and general parallel programming model, similar to MPI and CUDA. The user specifies the number of tiles $T$ at compile time and the DDG generates $T$ graphs of $K$. The compiler then creates a native binary that executes $K$ with $T$ threads using OpenMP, generating the necessary traces.

Accelerator tiles (further detailed in Section~\ref{sec:accel}) can be invoked via an API of common accelerated functions, e.g.\ \texttt{SGEMM}. The DDG captures the accelerator call and the DTG records the relevant parameters, e.g.\ matrix dimensions. During simulation, the accelerator node in the DDG is matched with the trace parameters and the accelerator model is invoked.
Sections~\ref{sec:case-studies} highlights examples of accelerator use.

\subsection{Inter-Tile Communication \label{sec:inter-tile-comm}}
Tiles operate alongside each other, each being called upon by the Interleaver (Figure~\ref{fig:digester}) to take a single-cycle step. Tiles can communicate through a traditional shared memory hierarchy, in which memory instructions (i.e.\ loads and stores) are dispatched to a memory model (discussed in Section~\ref{sec:memhier}). 

Two tiles can additionally communicate with each other through generic \emph{messages}, which can be stored in internal tile buffers.
This is realized through a simple message passing API (i.e.\ \texttt{send}, \texttt{recv}). The Interleaver buffers all \texttt{send} instructions issued. When the receiving tile issues a \texttt{recv} instruction, the Interleaver matches it with the buffered message. This model is simple, generic, and can be used to build more complicated and specialized inter-tile communication models. Section~\ref{sec:DAE-case-study} discusses how these features can be used to implement a Decoupled Access/Execute system~\cite{dae}.

\section{Fast Abstract Tile Models in \projname{}}
\label{sec:cpu}
As previously described, \projname{} can simulate various tile models that estimate the performance and power costs of a region of LLVM IR. Analysis of the LLVM IR dependence graphs can be shaped to accurately reflect the resource constraints of different tile design choices. This section describes the modeling of different execution scenarios that correspond to microarchitectural resource limits for different tile models.

\subsection{Microarchitectural Resource Limits \label{sec:resource_limits}}
In order to be instantiated, a core tile model requires several microarchitectural resource parameters, such as issue width, RoB size, LSQ size, and the number of functional units. Based on these limits, MosaicSim manages resources to accurately model in-order, out-of-order, and accelerator tiles. 

{\bf Issue Width:} \projname{} models a superscalar issue width $W$ by maintaining a count of issued instructions and ensuring that no more than $W$ instructions can issue each cycle. 

{\bf ROB:} To model an ROB, \projname{} creates IDs for all instructions
that are assigned at DBB creation time. \projname{} maintains a sliding instruction window (starts with ID 0 and spans the instruction window size) that only allows instructions with IDs within the window to issue. When the oldest issued instruction completes, \projname{} slides the instruction window forward to issue a younger instruction.

{\bf LSQ Size:} To model the LSQ, \projname{} uses the MAO (described in Section~\ref{sec:graphstomodels}) to track loads and stores and ensure that instructions cannot issue if the MAO is full. Memory operations free up space on the MAO upon completion.  

{\bf Live DBB Limits:}
\projname{} provides the option of limiting the number of live DBBs that can run concurrently for each basic block. This limit mimics restricting how many replicated circuits for a loop body appear in a hardware accelerator. Entire DBBs (and their instructions) cannot be launched if the live DBB limit for their basic block has been reached.

{\bf Functional Unit Limits:} 
\projname{} can limit the number of available functional units for each instruction type. It maintains a count of all issued, incomplete instructions and the functional units they utilize. There must be an available functional unit in order to issue an instruction. When instructions complete, they free up the functional units they occupied. 



\subsection{Instruction Costs} Individual instructions in \projname{} have both a latency cost (cycles) and energy cost (Joules). These costs can be pre-determined (computation instructions) or dynamic (memory operations). After an instruction with a fixed cost is issued, \projname{} ensures that it does not complete until its global cycle count has progressed through the latency of the instruction. The fixed energy cost of the instruction is then added to a running total. For instructions with a dynamic cost (e.g.\ a memory instruction), cost values are determined by querying the memory hierarchy and are subject to factors such as  memory contention and cache misses (detailed in Section~\ref{sec:memhier}).

\subsection{Speculation \label{sec:speculation}}

\projname{} is designed to flexibly explore several opportunities for speculation. \projname{} models control-flow speculation by adding a misprediction latency whenever a modeled branch predictor contradicts the pre-determined control flow path provided by the DTG\footnote{\projname{} currently supports static branch prediction in addition to perfect branch prediction. This is useful for early-stage modeling, e.g.\ obtaining upper bounds. However, future work will support more realistic dynamic branch predictors.}. By default, \projname{} must wait until it encounters the terminator node of a basic block before launching a new DBB. However, with speculation, the next DBB can be launched immediately, which makes instructions in the newly launched DBB eligible to be issued. Instructions in a mispredicted path are never executed, as is similar with other instrumentation-based or direct-execution simulators (e.g. Sniper~\cite{sniper} or ZSim~\cite{zsim}).


\projname{} also leverages information from the DTG to provide an option for perfect \textit{memory address alias speculation}. Since the trace holds information on \textit{all} addresses for all instructions before starting the simulation, \projname{} knows if any pair of accesses have aliasing addresses ahead of time. Hence, it can ``perfectly anticipate'' aliasing occurrences and potentially issue memory instructions in the presence of unissued, older instructions with unresolved memory addresses. 

\section{Accelerator Simulation}
\label{sec:accel}


As shown in Figure~\ref{fig:digester}, \projname{} supports the simulation of heterogeneous SoCs comprised of CPUs and accelerators. \projname{} offers two styles of accelerator simulation for design progressions from high-level to detailed.

{\bf Pre-RTL Accelerator Modeling: } Early in the design process, pre-RTL accelerator modeling can help determine which accelerators are useful without their RTL designs. For this purpose, \projname{} can model accelerators using the same graph-based approach as previously described for CPUs, but with different hardware resource constraints. Fixed-function accelerator models provision hardware resources based on application-specific factors (e.g.\ loop unroll length and parallelism opportunities). \projname{} provides knobs to specify the number of active DBBs per basic block (i.e.\ hardware-supported loop unrolling), number of functional units, etc. In addition, one can use \projname{} to explore the relaxation of hardware constraints, such as RoB size and instruction window. Rather than targeting a specific hardware implementation, these features enable a high-level exploration of the extent to which an application can benefit from hardware acceleration.

{\bf RTL Accelerator Modeling:}	 Later in the design process, \projname{} allows a high-level accelerator model to be replaced by a more detailed one based on an actual RTL implementation of the accelerator. This is essentially a substitution of one (or several) of the tile models depicted in Figure~\ref{fig:digester}.

\subsection{Accelerator Invocation}
When \projname{} invokes an accelerator, the Interleaver queries the accelerator tile for latency and resource usage information.  For graph-based accelerator modeling, this invocation is similar to that of CPU models.  

For detailed RTL accelerator modeling, \projname{} provides an interface tailored to such evaluations. The Interleaver runs a C++ performance model of the accelerator tile, which takes as input two sets of parameters: (1)~a standard set of generic system parameters, e.g.\ technology node, maximum memory bandwidth, number of accelerator instances to be invoked in the system; and (2)~a set of accelerator configuration parameters, e.g.\ number of inputs, input and output sizes. 
When queried, the accelerator tile model returns to the Interleaver a set of performance estimates, e.g.\ clock cycles, bytes of memory accessed, and average power consumption. This accelerator data is then included in the final performance results reported by \projname{}. Individual accelerator tiles can be implemented in various ways as long as they abide by the Interleaver's interface. We next describe our primary methodology for generating the accelerator tile models.

%

\subsection{RTL Accelerator Model Design}
\label{sec:acceldesign}
The RTL accelerator modeling approach focuses on accelerators with predictable memory access patterns, i.e. independent accesses. However, the overall \projname{} approach is more general and can support tile models with any access patterns.

\begin{figure}[!t]
\centering  
\includegraphics[width=\linewidth]{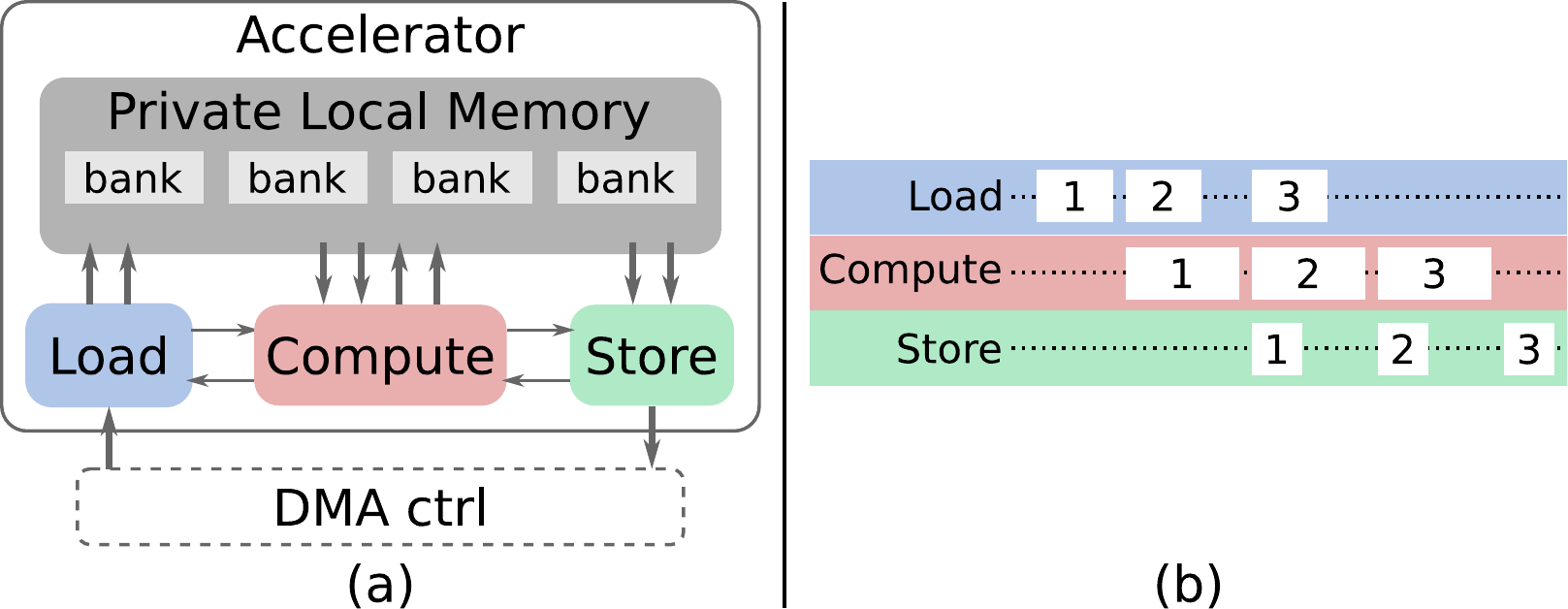}
\caption{(a) The accelerator's modules operate in a pipeline with a multi-port, multi-bank private local memory. (b) Computation and communication overlap during the accelerator execution.}
\label{fig:acc-scheme}
\end{figure}

In this modeling approach, accelerators are designed by leveraging the accelerator design flow of the open-source ESP project~\cite{esp_release, giri_date20, piccolboni_hpec17}, which eases the design effort by using templates to automatically generate most of the accelerator source code. Accelerators are first designed in SystemC and then sent through Cadence's Stratus High-Level Synthesis (HLS) tool to produce an RTL implementation. This methodology is applicable to other languages and tools, for instance ESP provides accelerator design flows also in C/C++, Keras TensorFlow, Pytorch and more.

The accelerators generated through this approach have a pipelined datapath crafted to mask the communication time as much as possible. \figurename~\ref{fig:acc-scheme} presents an accelerator with three concurrent modules: a load process to load data from memory, one or more computation processes, and a store process to send data to memory. The modules communicate through a private local memory, which is a circular or double buffer to enable pipelining of computation and DMA transfers.

{\bf Communication Model:}
In accelerator tile model development, validation of the communication and computation performance characteristics with respect to the expected hardware is key. Both the SystemC and the HLS-generated RTL implementation of an accelerator can be verified in simulation with a SystemC testbench. We augmented the testbench infrastructure of the ESP accelerators with a memory model that accounts for access latency, bandwidth, interconnect bit-width, and average NoC hops between accelerator and memory interfaces (for NoC-based SoCs). These parameters can be tuned to match a target SoC. With this kind of communication model, a designer can focus on the accelerator design without losing sight of its interaction with the rest of the system.

Accelerators can interact with the memory hierarchy in many ways~\cite{giri_ieeemicro18, shao_micro16}. This work models accelerators as non-coherent; they communicate directly with main memory, bypassing the memory hierarchy. This is common for loosely-coupled accelerators that execute coarse-grained tasks.

{\bf Performance Model:}
\projname{} has a generic performance model for loosely-coupled, reconfigurable, fixed-function accelerators. The model abstracts an accelerator as a set of concurrent modules, where each module executes one or more loops multiple times. The model can also invoke accelerators in parallel and, given a maximum memory bandwidth, scale execution time and average power consumption accordingly. 

The performance model of a specific accelerator employs the generic model by providing the following four arguments: (1)~the number of processes; (2)~the number of loops per process, which describes the accelerator structure; (3)~the total latency of all internal loops; and (4)~the number of iterations of each loop, which is function of the configuration parameters of the accelerator invocation.

The designer needs to provide the average power consumption of the accelerator, which can be measured by logic synthesis tools based on the switching activity recorded during RTL simulation. Finally, the accelerator designer should provide an expression to calculate the number of bytes transferred to/from memory as a function of the accelerator configuration.

{\bf Accelerator Instrumentation:}
The generic performance model requires the latency of one iteration of the core loops in each module as input. These are the internal loops, whose latencies do not depend on the accelerator configuration (e.g. input size). To aid the collection of these latencies, we augmented the \textit{ESP accelerator templates} with instrumentation features, so that the accelerator designer can instrument the accelerator to collect the required cycle-accurate latencies.

The instrumentation adds an array of signals to each module and an additional concurrent process, the \textit{collector}, to collect and process all instrumentation signals. Each signal is toggled at every iteration of the corresponding loop. The \textit{collector} measures the toggling latency and communicates the results to the testbench, which ultimately dumps them to a file. 


{\bf Design Space Exploration:}
HLS allows for seamless generation and evaluation of multiple RTL implementations of an accelerator given a single high-level SystemC specification. The SoC designer can then choose which specific design point(s) to instantiate as well as how many copies of the same accelerator should be present. The very fast system simulation of accelerators with \projname{} can greatly help this design-time decision process.


\section{Memory Hierarchy} 
\label{sec:memhier}


\projname{} simulates a memory hierarchy that includes caches: both private and shared, and support for two different DRAM models: an in-house model named SimpleDRAM, and the widely-used DRAMSim2~\cite{dramsim2}. 

\subsection{Cache Model} 
\projname{}'s cache model can be utilized as a per-core private cache or shared cache. Both are independently configurable for size, cache line size, associativity, and access latency. \projname{} is a timing simulator and therefore need not hold actual data in the caches; the address tags suffice. 

The cache hierarchy is conventionally write back, write allocate, and fully-inclusive. Each core tile model maintains a cache queue ordered with respect to the cache hierarchy. Memory requests are initially sent to the L1 cache at the front of the queue and forwarded to the next cache when necessary (e.g.\ cache misses or writeback of dirty data). At the end of the queue, the LLC forwards requests to the DRAM model (described in Section~\ref{sec:DRAM}). 

The cache model includes a prefetcher that detects streaming patterns of memory accesses. It simply tracks memory requests to see if there exists a chain of accesses that are $k$ words apart. If so, a number of additional requests are generated by the cache for subsequent cachelines in anticipation of future memory instructions accessing those cachelines. The number of cachelines prefetched and the address distance from the instruction triggering the prefetches can be configured. \projname{}'s memory hierarchy model provides a flexible and straightforward interface to implement more complex or specialized prefetchers as well.

To coalesce memory requests, caches can utilize an MSHR whose size can be configured. When a cache receives a request, it checks the MSHR to see if there exists a pending request to the same cacheline. If so, it saves the request on the MSHR. When the pending request is served, the MSHR notifies all requests waiting on that cacheline. 

Precise modeling of NoCs, consistency, and coherence are currently not implemented in MosaicSim, as this level of detail is not required by our early-stage modeling. However, future work aims to provide their support. With MosaicSim’s modular design, ports can be added to the abstract tile model to create a message module in order to model NoCs and the necessary communication for coherence and consistency. A directory protocol can easily be implemented by treating the Interleaver as the directory and allowing it to communicate with the caches.

\subsection{DRAM Model \label{sec:DRAM}}
\projname{} supports two DRAM models: an in-house model called SimpleDRAM and the widely-used DRAMSim2~\cite{dramsim2}. SimpleDRAM ensures that all DRAM requests abide by a minimum latency and maximum bandwidth. Every DRAM request is inserted into a priority queue ordered by minimum request completion time (current cycles plus minimum latency). SimpleDRAM enforces the maximum bandwidth limit in epochs. Every cycle, it attempts to return as many requests as possible that have served the minimum latency. Once the number of requests returned in that epoch has exhausted the maximum bandwidth, SimpleDRAM cannot return requests until the next epoch, but it can continue receiving new requests. SimpleDRAM thus models memory bandwidth contention and throttling due to bandwidth limits. 

SimpleDRAM is the default model, but \projname{} can be configured to use DRAMsim2 for cycle-accurate DRAM modeling, albeit this model executes slower has a larger memory footprint during simulation than SimpleDRAM.
\section{Evaluation \label{sec:evaluation}}

\setlength{\tabcolsep}{0pt}
\begin{table}[t]
\centering
\small
\caption{Evaluation system details Intel Xeon E5-2667 v3}
\label{tab:machine}
\begin{tabularx}{\columnwidth}{l Z}
  \toprule
  Sockets, Cores & 2 sockets, 8 cores each \\
  Node Technology and Frequency & 22nm, 3200 MHz \\
  L1-I and L1-D & 32KB private / 8-way  \\
  L2 & 2MB private / 8-way \\
  LLC & 20MB shared / 20-way \\
  DRAM & 128GB DDR4 @ 68GB/s \\
  \bottomrule
\end{tabularx}
\end{table}

We make use of our hardware-software toolchain to evaluate \projname{} on a variety of benchmarks. The simulator relies on the compiler to generate the DDG and the DTG to instrument the code and generate memory and control flow path traces. 
\projname{} utilizes the front-end tools in the stack to quickly and accurately simulate heterogeneous and hardware-software co-design systems.

\subsection{Accuracy \label{sec:accuracy}}

In order to measure \projname{}'s ability to accurately capture and characterize application trends, we perform two evaluations. First, we utilize the Parboil benchmark suite~\cite{parboil} to evaluate core and memory hierarchy models. We validate \projname{} by running benchmarks on the Intel Xeon E5-2667 v3 processor (features detailed in Table~\ref{tab:machine}) and collecting measurements of our real machine using Intel VTune Amplifier~\cite{vtune}. 
By using VTune's function-level filtering to isolate profiling information for the kernel, we obtain cycle and instruction counts to compare against \projname{} performance estimates. Second, we evaluate our accelerator tile models against both RTL simulation as well as FPGA execution.

\begin{figure*}[t]
    \centering
    \begin{minipage}[t]{0.49\textwidth}
        \centering
        \includegraphics[width=\linewidth]{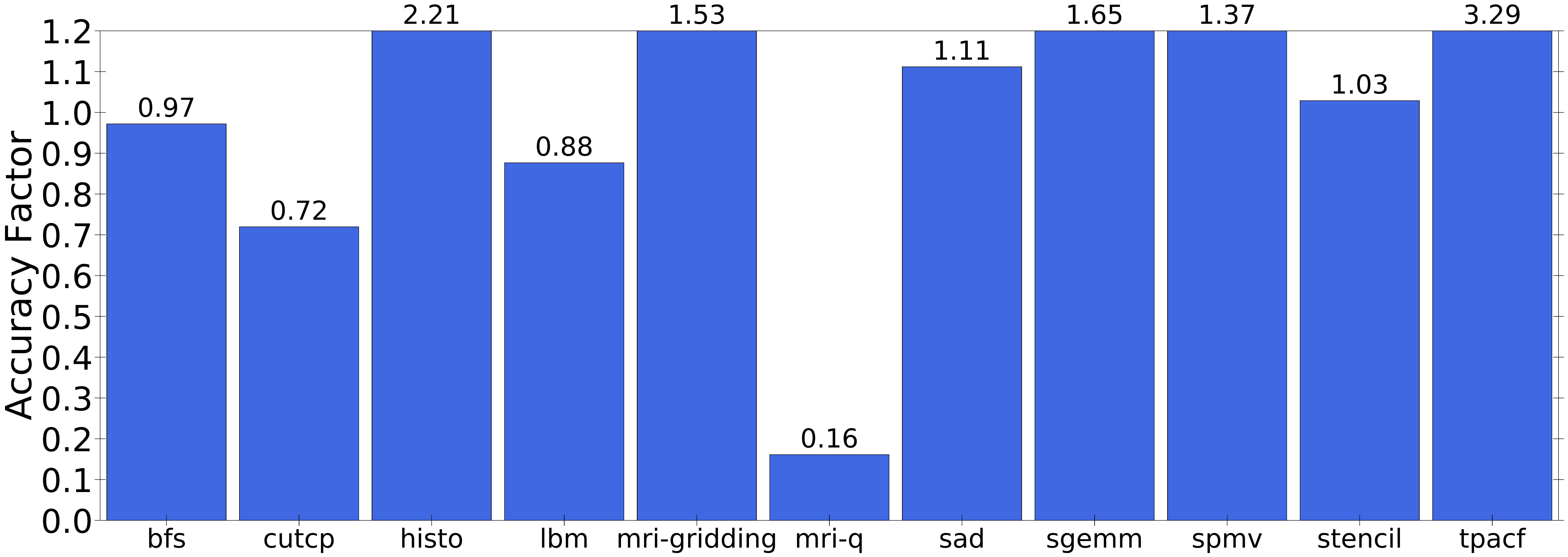}
        \caption{Despite inaccuracies in individual benchmarks due to ISA differences, \projname{} achieves a geomean runtime accuracy of 1.099$\times$ against an x86 machine.}
        \label{fig:runtime}
    \end{minipage}
    \hfill
    \begin{minipage}[t]{0.49\textwidth}
        \centering
        \includegraphics[width=\linewidth]{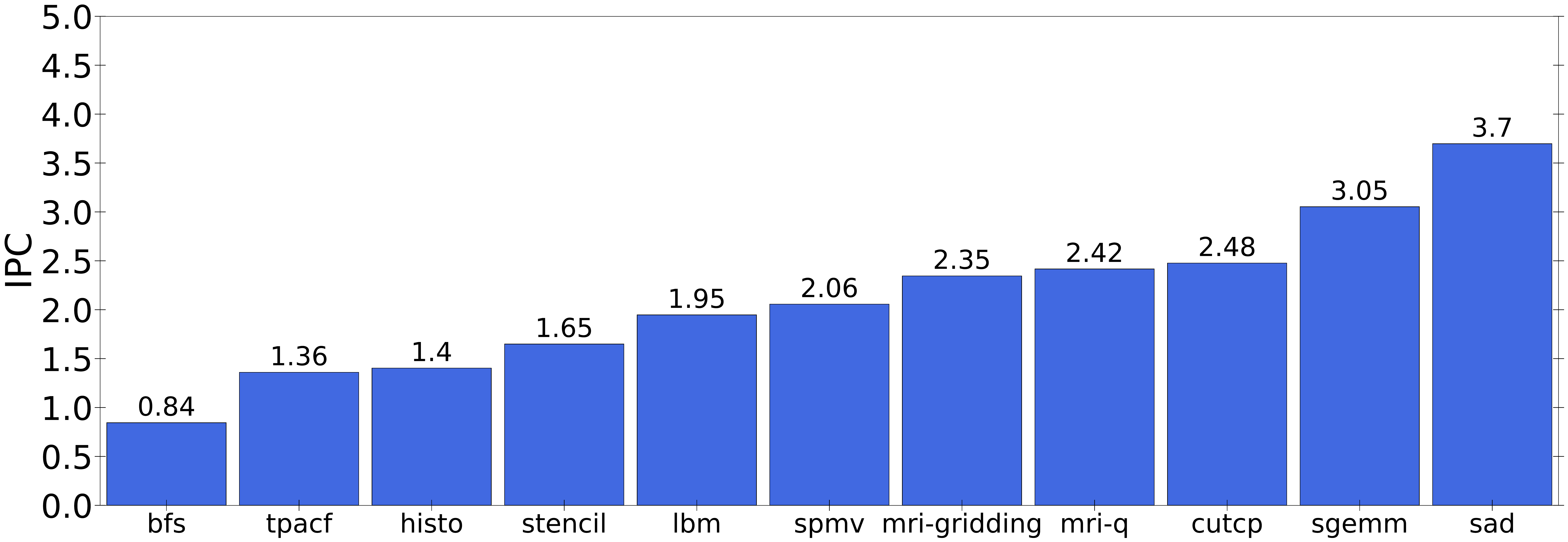}
        \caption{MosaicSim accurately characterizes applications with IPC measurements (lower implies more memory-bound while higher implies compute bound).}
        \label{fig:ipc}
    \end{minipage}
\end{figure*}

{\bf Application Characterization:} Figure~\ref{fig:runtime} displays the accuracy of \projname{}'s runtime estimates compared to the measured performance of real hardware. \projname{} achieves a geomean accuracy (simulated cycles/real cycles) of 1.099$\times$. 
Accuracy discrepancies arise from MosaicSim being ISA-agnostic; it cannot perfectly capture cases where LLVM IR instructions do not have a direct, 1-to-1 mapping to actual ISA instructions. For example, LLVM IR requires two instructions for loading from an address offset: \texttt{load} and \texttt{getelementptr}, while the x86 ISA can perform this with one instruction: \texttt{MOV}. Additionally, a direct comparison of LLVM IR simulation against a native ISA must take into account compiler optimizations applied when producing the binary from the IR (e.g.\ we have found that using \texttt{-O3} and loop unrolling produces a more accurate comparison to an x86 instruction counts). Thus, we expect precise IPC and timing models to be noisy when compared to the execution of a concrete ISA. Figure~\ref{fig:runtime} demonstrates this behavior on the Parboil suite: although the geomean accuracy is high, individual benchmark measurements can be inaccurate. We have found that fine-grained tuning of LLVM IR simulation for concrete ISAs, e.g.\ simulating pairs of \texttt{load} and \texttt{getelementptr} as one instruction for x86, can increase accuracy. However, \projname{} aims to be ISA-agnostic and therefore focuses more on characterization rather than on raw cycle accuracies.

Due to the extra abstraction layer of LLVM IR, it is difficult to perform raw IPC comparisons without tuning to a specific ISA. However, we can use the IPCs that \projname{} reports to characterize kernels as memory or compute-bound. A lower IPC indicates that a kernel is memory-bound while a higher IPC indicates being compute-bound. These results, e.g.\ \texttt{BFS} being memory-bound and \texttt{SGEMM} being compute-bound, match previous characterizations of these common benchmarks~\cite{desc, parboil}.

{\bf Scaling Trends:} In order to evaluate \projname{}'s ability to capture scaling trends, we measure both simulated and real hardware performance for \{1, 2, 4, 8\} thread(s). We then normalize all performance numbers to those with a single thread and evaluate how benchmark speedups scale with an increasing number of threads.  

\begin{figure*}[t]
    \centering
    \begin{minipage}[t]{0.32\textwidth}
        \centering
        \includegraphics[width=\linewidth]{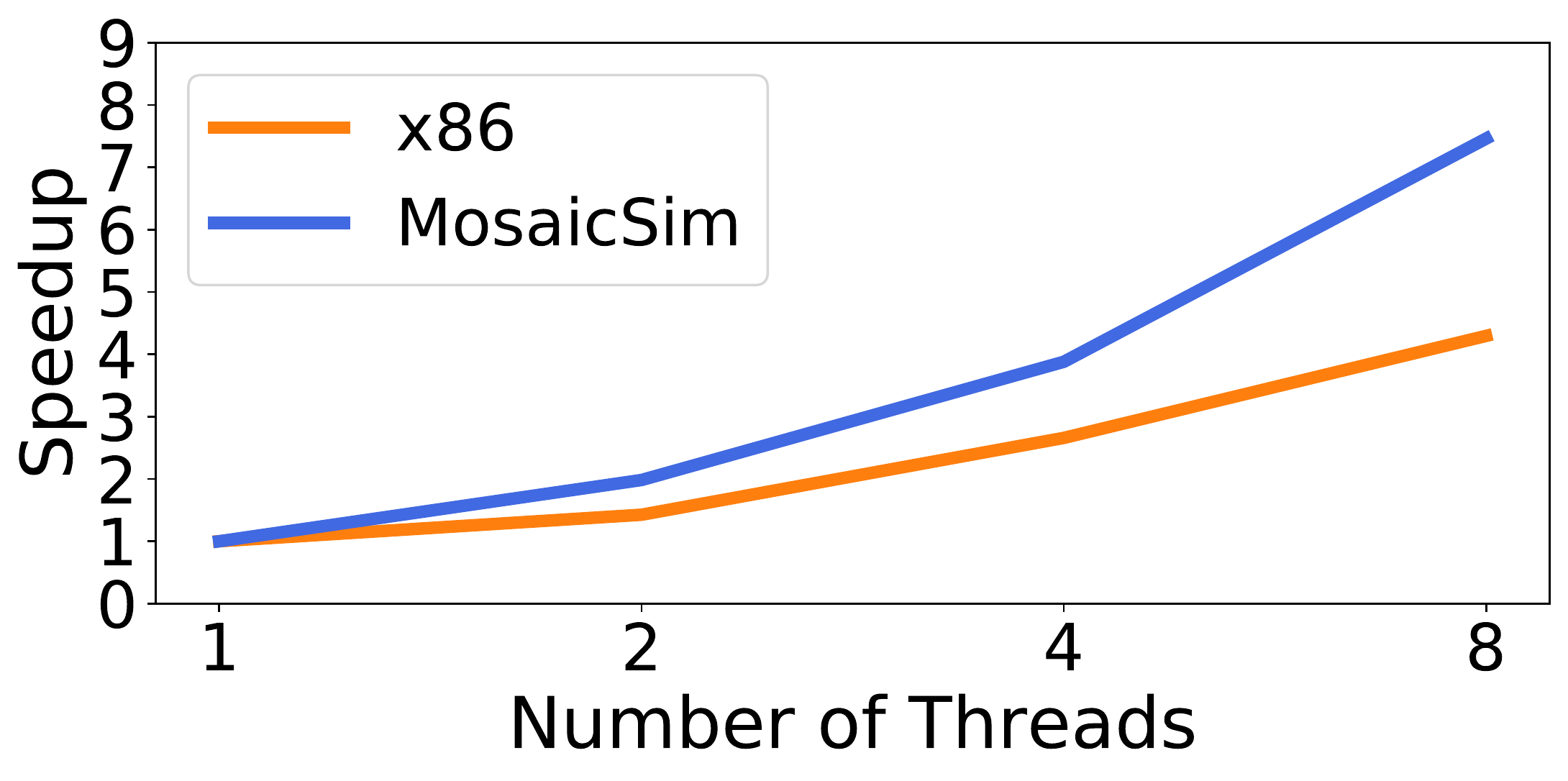}
        \caption{\texttt{BFS} Scaling Trends}
        \label{fig:bfs_scaling}
    \end{minipage}
    \hfill
    \begin{minipage}[t]{0.32\textwidth}
        \centering
        \includegraphics[width=\linewidth]{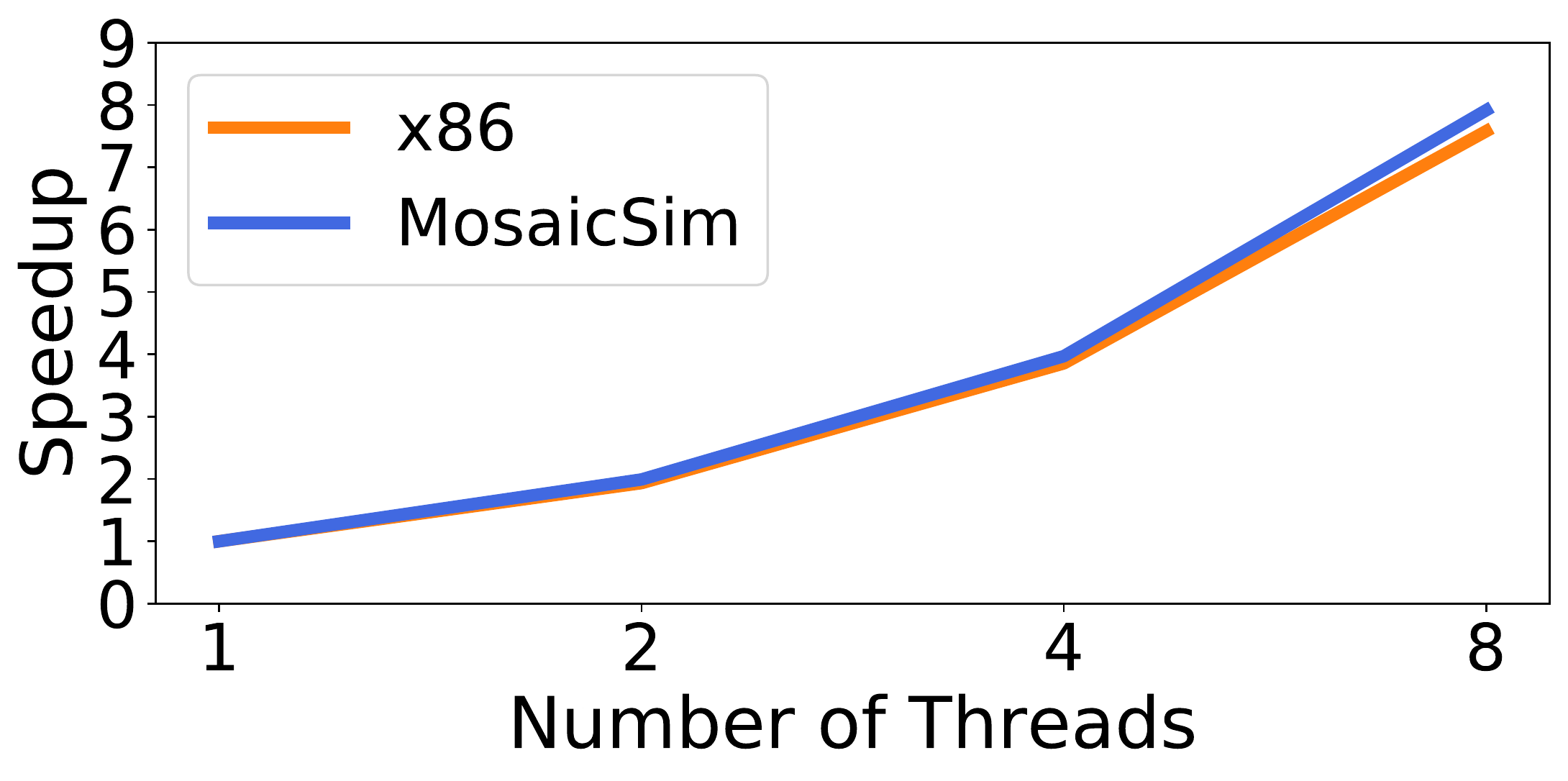}
        \caption{\texttt{SGEMM} Scaling Trends}
        \label{fig:sgemm_scaling}
    \end{minipage}
    \hfill
    \begin{minipage}[t]{0.32\textwidth}
        \centering
        \includegraphics[width=\linewidth]{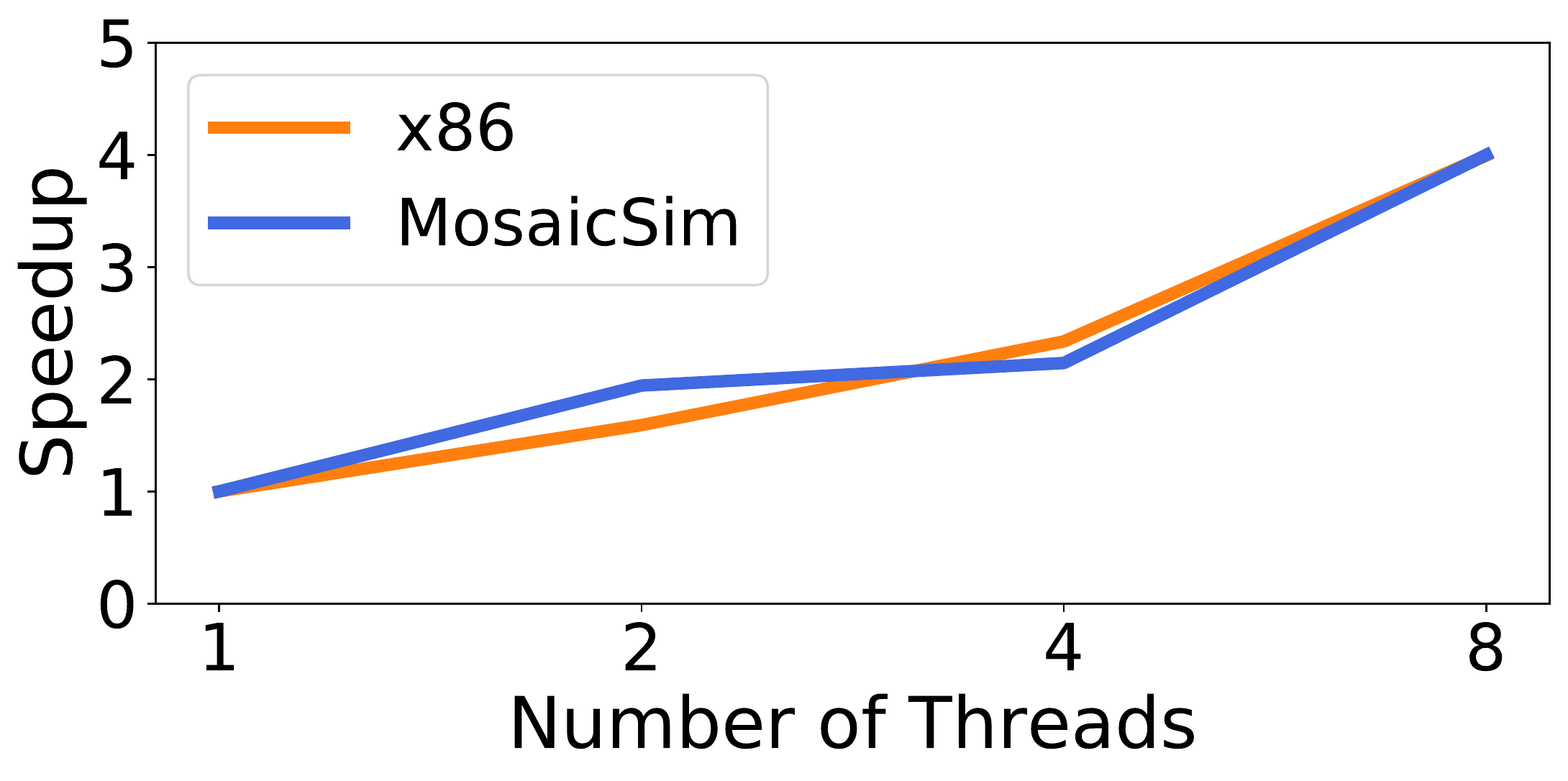}
        \caption{\texttt{SPMV} Scaling Trends}
        \label{fig:spmv_scaling}
    \end{minipage}
\end{figure*}

Figures~\ref{fig:bfs_scaling} -~\ref{fig:spmv_scaling} highlight the comparison of scaling trends for three well-studied benchmarks with different performance bottlenecks: \texttt{BFS} (latency-bound), \texttt{SGEMM} (compute-bound), and \texttt{SPMV} (bandwidth-bound), respectively. \projname{} nearly perfectly captures the linear scaling trend of \texttt{SGEMM} as the kernel is compute-bound and exposes data-level parallelism. 
\texttt{SPMV} is bandwidth bound, i.e.\ memory accesses are occasionally throttled, and we accurately capture the resulting sublinear scaling trend here. \projname{} is not as accurate on the latency-bound \texttt{BFS} kernel due to the use of atomic read-modify-write instructions that are difficult to accurately model in the memory system (Section~\ref{sec:memhier}); future work aims to more accurately model these instructions.

Being ISA-agnostic, MosaicSim demonstrates usefulness as an early-stage tool goal with its ability to capture performance bottlenecks and characterizations. Scaling and IPC characterizations are accurate and in line with prior work. If a designer later requires runtime accuracy for a given ISA, it is possible to add fine-grained tunings for LLVM IR simulation to help account for ISA discrepancies.


\begin{figure*}[!t]
\centering  
\includegraphics[width=1\linewidth]{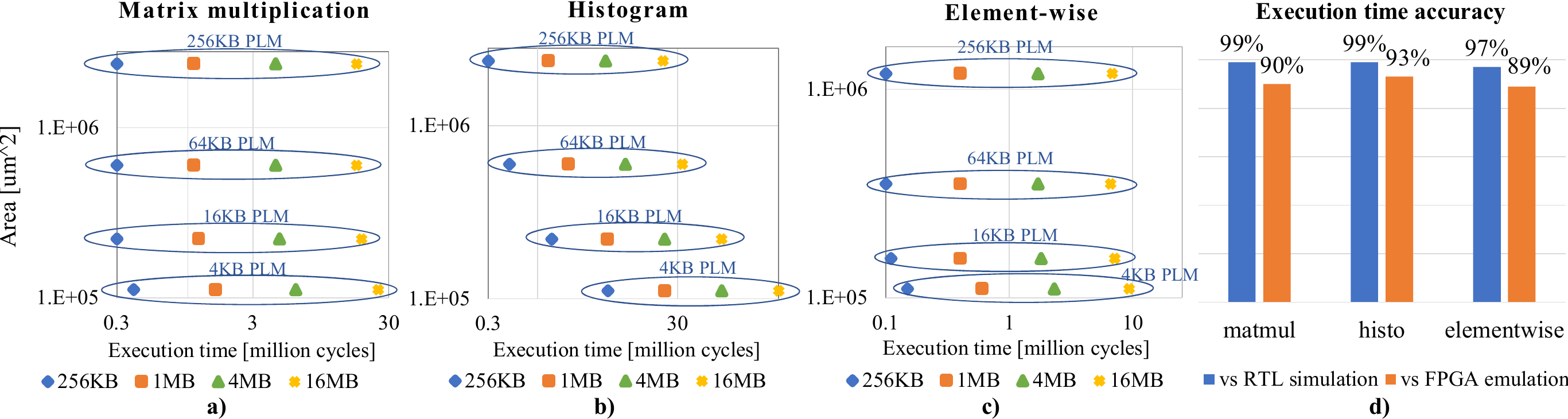}
\caption{a,b,c) Design space exploration for multiple workload sizes of 3 reconfigurable fixed-function accelerators. d) Accuracy of RTL-based performance models with respect to RTL simulation and full-system FPGA emulation.}
\label{fig:dse}
\end{figure*}

{\bf Accelerator Simulation:}
With the design flow described in Section~\ref{sec:accel} we created three fixed-function accelerators for matrix multiplication, saturating histogram, and element-wise arithmetic. These accelerators support any input size and any number of inputs per invocation. Using the ESP platform~\cite{giri_ieeemicro18, giri_nocs18}, we deployed the accelerators on a Xilinx Ultrascale+ FPGA as part of a many-accelerator SoC capable of running Linux. Therefore, we were able to validate the accelerators both with RTL simulation and on FPGA.
With HLS we generated multiple design points for each SystemC specification of the accelerators. Figures~\ref{fig:dse}a-c shows the execution time and area of four design points (with varying PLM size) and four workload sizes. Each design point is a distinct RTL implementation of the accelerator, whose performance model can be invoked by \projname{}.

\figurename~\ref{fig:dse}d shows the execution time accuracy of the models against RTL simulation of the accelerator and against full system FPGA emulation. The accuracy of each accelerator is the average of its accuracies for all the data points and workload sizes in Figures~\ref{fig:dse}a-c. The average accuracy with respect to RTL simulation is between $97\%$ and $100\%$, proving that our back-annotated generic performance model captures precisely the behavior of the accelerators. Furthermore, the models exhibit high accuracy ($>89\%$) when compared to a full SoC running on FPGA, validating the communication model that we added to the \textit{ESP accelerator templates}.

Recent literature shows that for medium to large workloads the overhead of the accelerator invocation through a Linux device driver is negligible~\cite{giri_ieeemicro18,shao_micro16}. We confirmed these results by measuring the overhead of invoking the accelerators, by invoking them on trivial workloads. We found that the overhead is consistently below $1\%$ of the execution time for the design points in \figurename~\ref{fig:dse}.

These RTL-based accelerator performance models do not actually execute the workloads and therefore take nearly no time to execute. They are several orders of magnitude faster than both RTL simulation and \projname{}'s pre-RTL accelerator modeling. In fact, these performance models are even faster than FPGA execution of the workloads they model.


\subsection{Using \projname{}}
This section describes practical details of using  \projname{} as an early-stage design tool for hardware-software co-design.

{\bf Designer Effort:} \projname{} provides a comprehensive set of both core and system configuration files that include a number of reconfigurable parameters (e.g. ROB size, issue-width, memory hierarchy details, etc.). These are straightforward to modify or extend, providing minimal designer effort. 

{\bf Simulation Speed:} \projname{} has a competitive simulation speed, achieving a single-threaded speed of up to 0.47 MIPs. This is comparable to that of Sniper~\cite{sniper_eval} (up to 0.45 MIPS) and is one order of magnitude better than gem5~\cite{gem5_riscv} (up to 0.053 MIPS). When the simulated system includes coarse-grained accelerator performance models (see Section~\ref{sec:accel}), the simulation speed is even higher, as many cycles of accelerator contributions are derived from a closed form equation (using  parameters obtained from a dynamic trace).

{\bf Storage Requirements:} As described in Section~\ref{sec:graphstomodels}, \projname{} requires both a DDG and memory control flow traces in order to run. The sizes of the DDG and control flow traces are typically less than 1 GB, thus we consider them negligible. However, the memory traces can be several GB large depending on the kernel. For example, in using the default datasets in Parboil, \texttt{BFS} takes 1.3 GB, \texttt{HISTO} takes 1.4 GB, and \texttt{SGEMM} takes 99 MB. While these traces can be large, they are necessary for accurate dynamic modeling of application behavior. \projname{} therefore aims to strike an appropriate balance between space efficiency and accuracy.

\section{Case Studies \label{sec:case-studies}}
In addition to across-the-board studies of simulation characteristics and accuracy, we provide three case studies that demonstrate the value of \projname{} to model complex heterogeneous systems and perform hardware-software co-design.

\subsection{DAE for Latency Tolerance \label{sec:DAE-case-study}}

The Decoupled Access/Execute (DAE) paradigm~\cite{dae} has been widely explored as a technique to tolerate memory latency by dividing a kernel into an \emph{access} slice and an \emph{execute} slice. The access slice performs \emph{all} memory accesses and all computation for an access, i.e.\ address computations and control flow statements where memory data is involved. Meanwhile, the execute slice performs value computation. 

The access slice performs loads and enqueues their data into a communication buffer between the access and execute slices. When the execute slice encounters a load, it simply reads the data from this buffer. Store instructions work conversely.
%
%
The key idea is that if the access slice can run ahead of the execute slice and produce all of the data required for computation, it can essentially act as a non-speculative ``perfect prefetcher''. The buffers in DAE are generally proposed in hardware implementations, leading to a \emph{heterogeneous} system, consisting of access and execute cores that execute their respective program slices concurrently.



Due to \projname{}'s support for heterogeneity, we can implement and evaluate DeSC~\cite{desc,descj,descpar}, a recently proposed DAE-based system. \projname{} allows us to evaluate DeSC on different (multi)core models (out-of-order and in-order), and perform area-equivalent design space exploration. 

{\bf Compiler and Simulator Support:} DAE program slicing can be implemented in the LLVM toolchain as a compiler pass. The pass first creates two copies of the kernel, one for access and one for execute. On the access slice, each memory instruction is augmented with a special function to either (1)~push to the buffer for loads or, (2)~replace a store value with a value from the buffer for stores. The execute slice is transformed similarly.

The DAE simulator support uses \projname{}'s inter-tile message-passing capabilities (Section~\ref{sec:inter-tile-comm}) to provide direct communication between the access and the execute cores. The load buffer is a \texttt{send} from the access slice and a \texttt{recv} from the compute slice. The store buffer is implemented conversely. Thus, the Interleaver processes these fine-grained inter-tile messages naturally. 
Additionally, the default core models were extended to include the structures described in \cite{desc}, i.e.\ communication queues, the terminal load buffer, the store address buffer, and the store value buffer.


\setlength{\tabcolsep}{0pt}
\begin{table}
\small
\caption{Parameters for DAE case-study.}
\label{tab:chip_params}
\begin{tabularx}{\columnwidth}{L{4cm} R{3cm} Z}
  \toprule
  \textbf{Microarchitecture\ Parameter}&  \textbf{Out-of-Order} & \textbf{In-Order} \\
  \midrule
  Issue Width & 4 & 1 \\
  Instruction Window/RoB/LSQ & 128/128/128 & 1 \\
  Frequency/Tech & 2GHz/22nm & 2GHz/22nm \\
  Area (mm$^2$)  &  8.44 & 1.01 \\  

  \bottomrule
  \end{tabularx}
\setlength{\tabcolsep}{6pt}

\setlength{\tabcolsep}{0pt}
\centering
\small
\label{tab:mem_hierarchy}
\begin{tabularx}{\columnwidth}{l Z}
  \toprule
  {\bf Memory Parameter} & {\bf Values} \\
  \midrule
  L1 & 32KB / 22nm node / 8-way / 1-cycle latency  \\
  L2 & 2MB / 22nm node / 8-way / 6-cycle latency \\
  DRAM &  DDR3L / 24GB/s BW / 200-cycle latency \\
  Comm.\ Buffer Sizes & 512 entries / 1-cycle latency \\

  \bottomrule

\end{tabularx}

\end{table}

\begin{figure}[!t]
\centering  
\includegraphics[width=\linewidth]{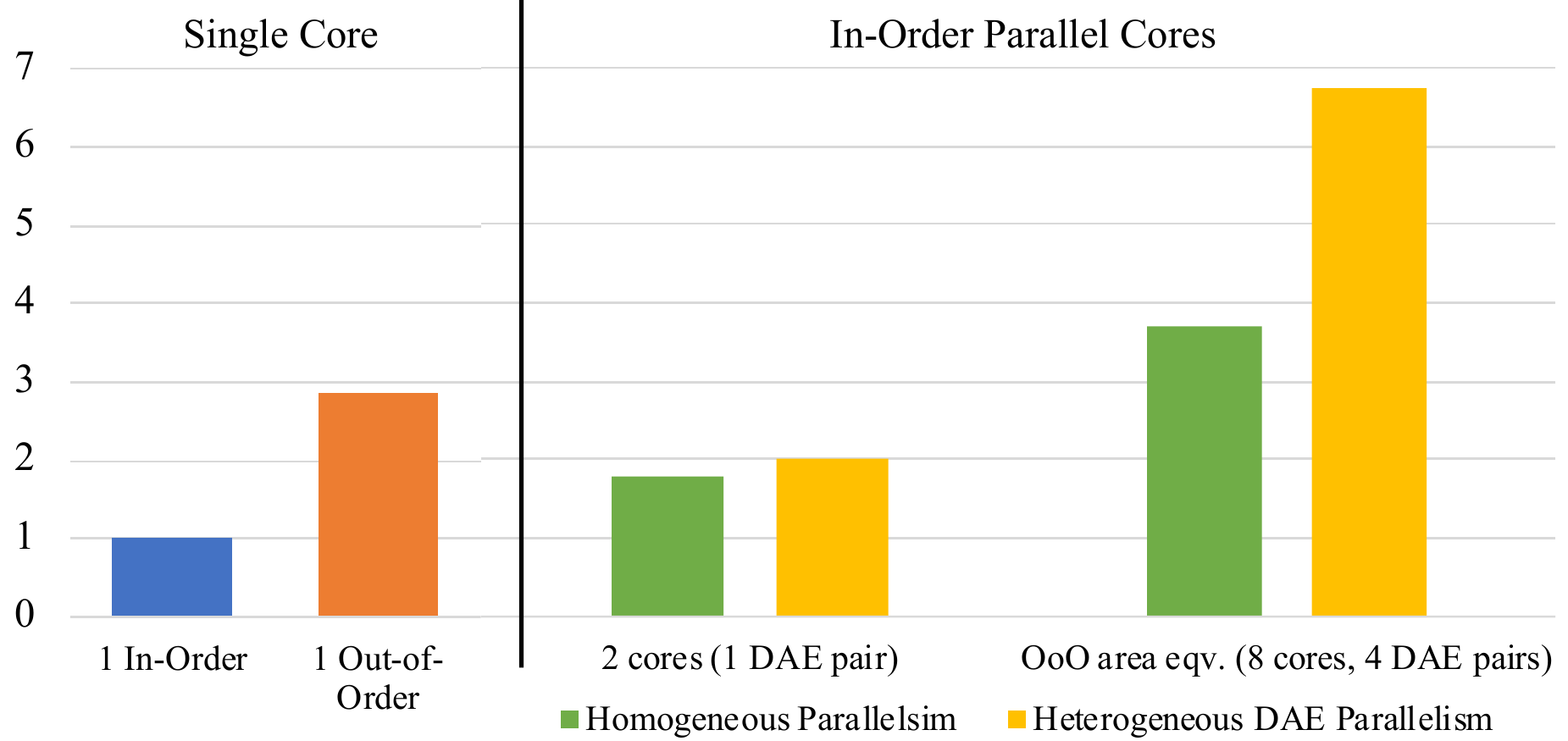}
\caption{Speedups of different systems on the graph projections kernel. In an equal-area comparison of 8 In-Order cores to 1 OoO core (right), DAE heterogeneity outperforms OoO by nearly 2$\times$, and is a promising approach for latency tolerance.}
\label{fig:dae-speedups}
\end{figure}

{\bf Results:} We evaluate our DAE implementation on the bipartite graph projection kernel, which has a wide set of use cases from recommendation systems \cite{bipartiteRec} to disease association prediction~\cite{bipartitemiRNA}. This application is memory latency bound; each pair of edges in the original bipartite graph updates a projection edge, which creates an irregular memory access.

We consider two base core models: in-order (InO) and out-of-order (OoO) (see Table~\ref{tab:chip_params}, area measurements are from McPAT~\cite{mcpat}). We augment the in-order model with DAE components to instantiate a parallel heterogeneous system where half the cores are access and the other half are execute.

Figure~\ref{fig:dae-speedups} highlights the results of this case study. We measure the performances of single-core, and homogeneous and heterogeneous parallel systems, and normalize them to that of a single InO core. As seen on the left, the OoO core, equipped with latency tolerance mechanisms, significantly outperforms the InO core. 
The right side presents scaling to 2 cores or 1 DAE pair and an OoO area-equivalent scaling to 8 cores or 4 DAE pairs. We see near-linear scaling for homogeneous parallelism (green bars), as a linear number of memory requests are issued in parallel.
Finally, we see that heterogeneous parallelism (yellow bars) yields the highest speedups (nearly 2$\times$) via asynchronous issuing of memory requests (proposed by modern DAE systems~\cite{desc}) and significant memory-level parallelism. 
%
Thus, \projname{} has enabled us to explore a heterogeneous system design as a promising approach for parallel, latency tolerant architectures.





\subsection{Alternating Sparse-Dense Phases \label{sec:sinkhorn}}
To further highlight \projname{}'s ability to simulate complex heterogeneous systems, we explore the architectural design space for applications which have both dense linear algebra (typically compute-bound) and sparse linear algebra (typically memory-bound). For example, Sinkhorn Distances~\cite{SinkhornDistances} is an algorithm for solving the optimal transportation problem and is used in computer vision~\cite{emd_computervision} and NLP~\cite{WMD}. The bottleneck of the application is split between a dense matrix multiplication (\texttt{SGEMM}) and an element-wise matrix operation where one operand is sparse and one is dense (\texttt{EWSD}).


\begin{figure}[!t]
\centering  
\includegraphics[width=\linewidth]{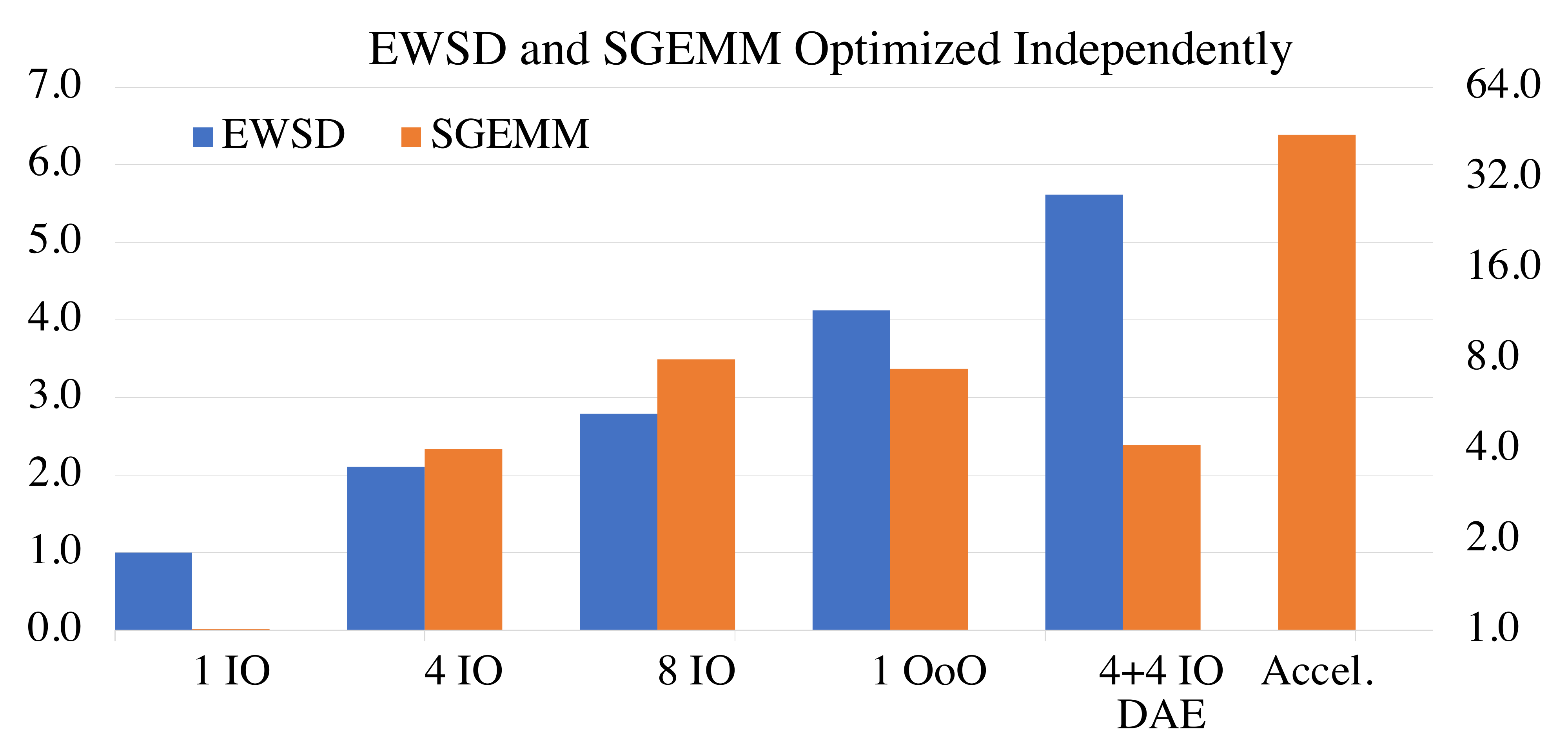}
\caption{Speedups of different systems on the \texttt{EWSD} (left axis) and \texttt{SGEMM} (right axis) kernels. \texttt{EWSD} benefits from latency tolerant
architectures, such as OoO and DAE systems. \texttt{SGEMM} benefits most from an accelerator. 
}
\label{fig:sinkhorn:ewsd_dense_separate}
\end{figure}

{\bf Architectures with Multiple Objectives:}
\begin{figure}[!t]
\centering  
\includegraphics[width=\linewidth]{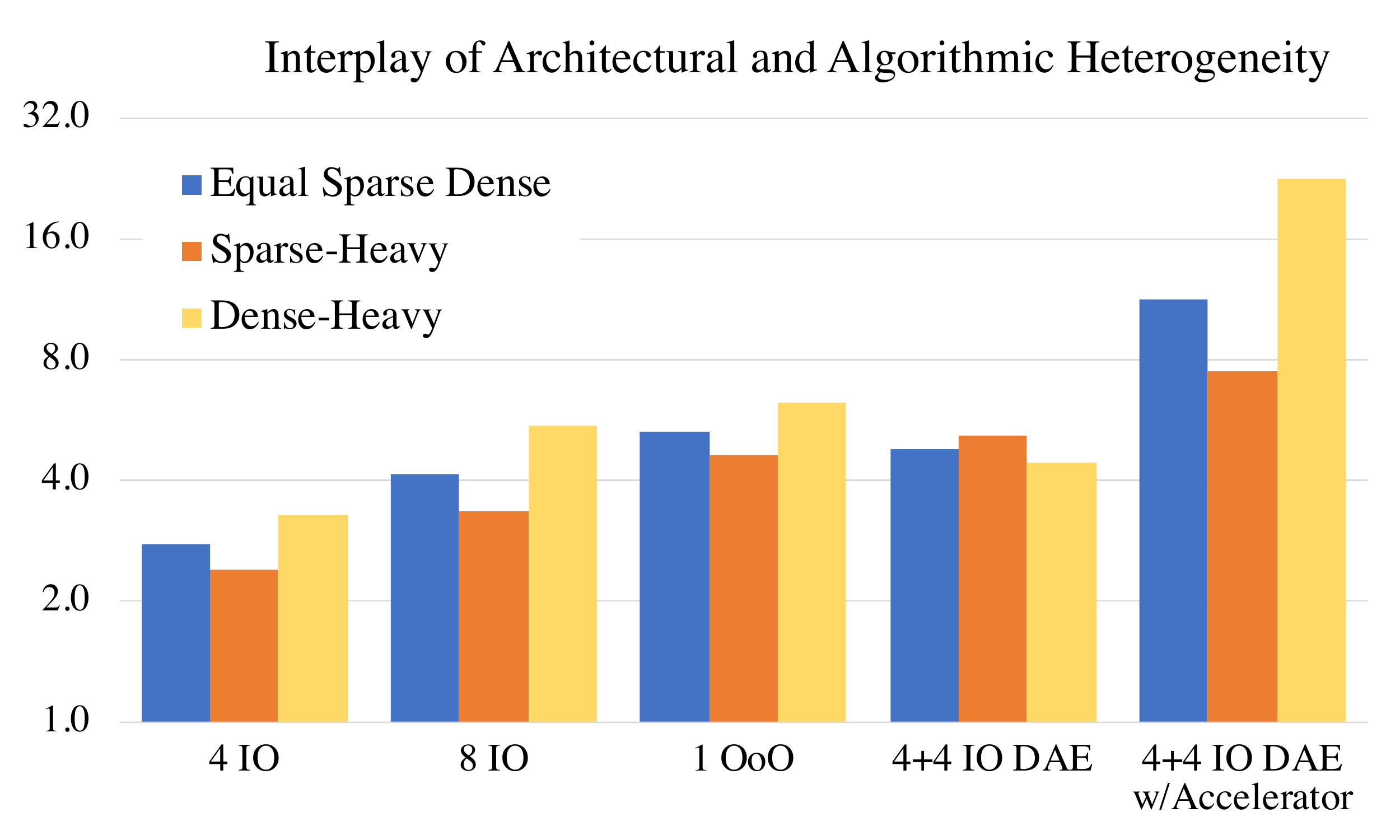}
\vspace{-2.5em}
\caption{A heterogeneous system executing a combined kernel of both dense (\texttt{SGEMM}) and sparse (\texttt{EWSD}), with one IO core being the baseline. The most heterogeneous system
has the best performance (DAE with an accelerator).}
\label{fig:sinkhorn:joint_ewsd_dense}
\end{figure}
To study the architectural design space for these types of applications, we start with constructing two microbenchmarks: \texttt{SGEMM} alone and \texttt{EWSD} alone. We simulate the runtime of each microbenchmark under various system configurations (see Tables~\ref{tab:chip_params}) and present the results in Figure~\ref{fig:sinkhorn:ewsd_dense_separate}. We use one InO core as the absolute baseline, as it is the simplest system. Since \texttt{SGEMM} is a compute-bound kernel, we consider the use of a fixed-function accelerator. Specifically, we use the matrix multiplication accelerator introduced in Section~\ref{sec:accuracy}.

The two microbenchmarks have different architectural performance landscapes. An optimal architecture for a kernel that combines them therefore needs to resolve conflicting demands. \texttt{SGEMM} sees large improvements from computation resources, as a fixed function accelerator for \texttt{SGEMM} provides nearly 45$\times$ speedup. Meanwhile, \texttt{EWSD} is memory bound and benefits from the latency tolerance available in a DAE architecture, which provides nearly a 6$\times$ speedup. 

{\bf Heterogeneous System Simulation:}
\projname{}'s main strength is simulation support for a complex heterogeneous system. To demonstrate this, we now construct a \emph{combined} benchmark that performs \texttt{SGEMM} and \texttt{EWSD} kernels serially. We then instantiated them with three different dataset sizes, where we varied the percentage of the total number of cycles spent in \texttt{SGEMM} versus \texttt{EWSD} based on their expected number of cycles on one InO core. This yielded a dense-heavy (75\% \texttt{SGEMM}, 25\% \texttt{EWSD}), a sparse-heavy (25\% \texttt{SGEMM}, 75\% \texttt{EWSD}), and an equally divided kernel. These combinations model  workloads found in real-world applications~\cite{SinkhornDistances, emd_computervision,WMD}.

Figure~\ref{fig:sinkhorn:joint_ewsd_dense} summarizes speedups of various architectures. Depending on the ratio of execution time for each of the two phases, the optimal architecture for the combined approach is non-trivial and requires the simulation of \emph{both} phases using a variety of tiles that make up a complex heterogeneous system. 
Our results show that in the absence of a specialized accelerator for the dense operation, the combined kernel would benefit most from 4 DAE pairs if the kernel is sparse-heavy and 1 OoO core if it is dense-heavy. With an accelerator however, 4 DAE pairs are the ideal choice for all cases. 
\projname{} allows the exploration of many combinations and configurations through its lightweight plug-and-play interface.


\subsection{Performance Modeling of TensorFlow Programs:}
To further demonstrate accelerator performance modeling with \projname{}, we present an example of simulation support for Keras TensorFlow programs. Keras~\cite{keras} is TensorFlow's high-level API for designing and training deep learning models. Applications of interest are therefore composed of multiple neural network kernels, e.g. convolution, matrix multiplication, pooling, etc. These kernels are computationally intensive and significantly contribute to the overall execution time of deep learning applications. Therefore, they are often deployed on accelerators.  
Thus, \projname{} can generate performance estimates of a Tensorflow application using accelerator performance models. 

To demonstrate this, we added a Keras TensorFlow API in the compiler to recognize Keras function names in the source code and map them to LLVM accelerator invocation calls when the application is compiled. These function calls are preserved as special instructions in \projname{}, where we add accelerator performance models for ESP accelerators~\cite{esp_release} according to the design flow described in Section~\ref{sec:acceldesign}. These accelerators provide kernel support for convolution, matrix multiplication, activation, pooling, etc. The accelerator invocation calls then appear in the instrumented LLVM that \projname{} operates on, so once the application is compiled and executed, the accelerator invocations are simulated whenever \projname{} encounters their function calls. We evaluate \projname{}'s TensorFlow application performance modeling with three deep neural network applications below.

 \texttt{ConvNet} is a type of convolutional neural network (CNN) application. CNNs are used to extract spatial, temporal and spatiotemporal relationship in data such as images, protein structure, language, and weather. The \texttt{ConvNet} algorithm contains an initial convolutional layer followed by a ReLu nonlinear layer that is regularized by batch normalization. This is followed by three residual blocks containing convolutional and residual layers. The final residual block is connected to a pooling layer and the model ends with a fully connected and activation layer that outputs a classification prediction. 

\texttt{GraphSage} combines graph and neural network algorithms and can be used as a recommendation system \cite{graphsage}. The objective of the algorithm is to sample graph data through a random walk and transform this data into a dense vector format that can be fed into a neural network architecture consisting of fully connected layers and ReLU layers. The algorithm mimics the continuous bag of words (CBOW) algorithm where instead of words, visited nodes are inserted into the input vector.

\texttt{RecSys} is a recommendation system modeled using neural networks. Training the algorithm takes as input individuals' preferences out of many available options, where the data is vectorized and fed into the model in batches. The neural network itself contains two sequential fully connected layers with ReLU nonlinear steps which are regularized with batch normalization and dropout methods. These layers are followed by a final fully connected layer which outputs new items that the model recommends.

\begin{figure}[!t]
\centering  
\includegraphics[width=\linewidth]{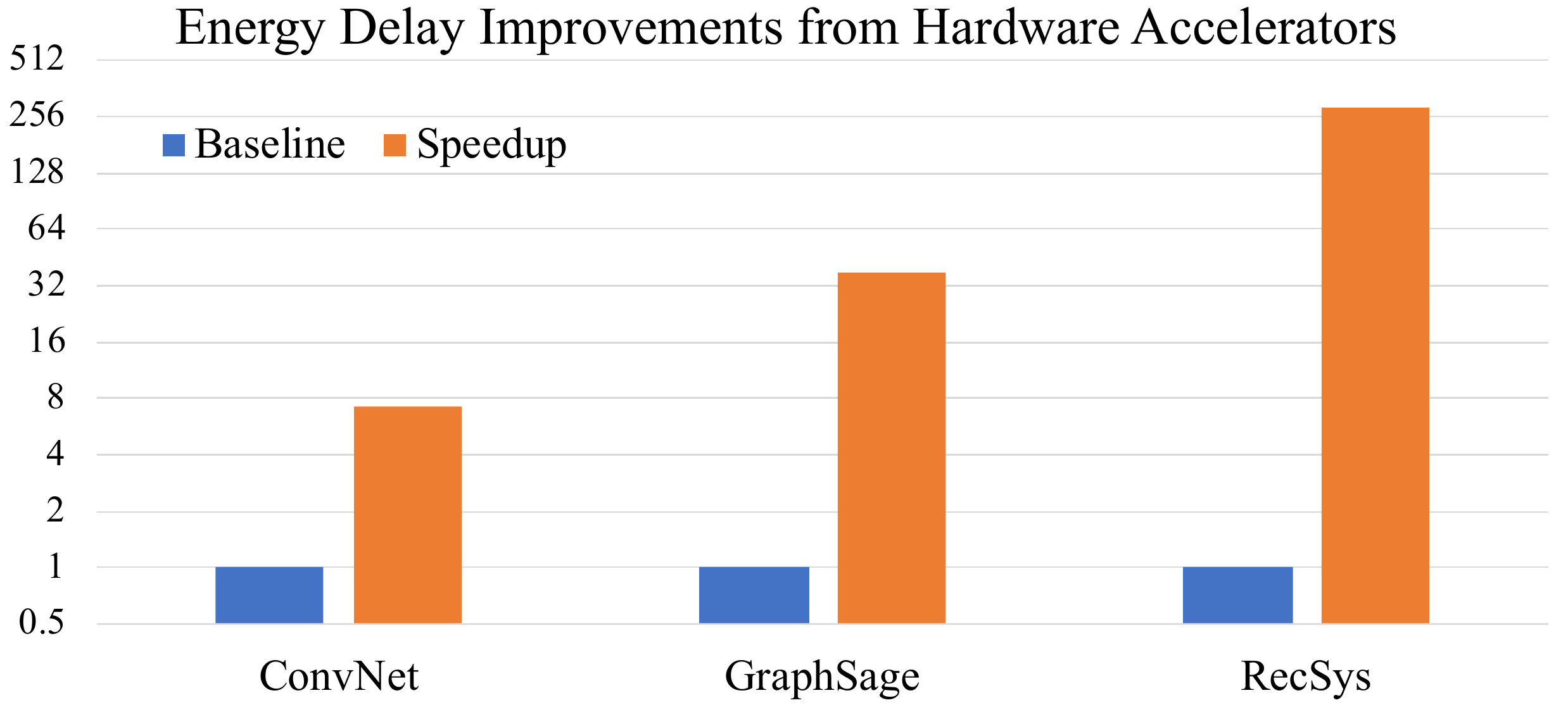}
\vspace{-2em}
\caption{Energy-delay improvement comparison between out-of-order cores and accelerator-oriented SoCs for three DNN applications (\texttt{ConvNet}, \texttt{GraphSage}, and \texttt{RecSys}).}
\label{fig:tf}
\end{figure}

We simulate and compare the performance of training of these three applications on two systems: an out-of-order server core with no accelerators and an SoC integrating 8 accelerators. We measure performance in energy-delay product, a metric which combines runtime and energy efficiency. Figure~\ref{fig:tf} highlights the comparisons, showing that \texttt{Convnet}, \texttt{GraphSage}, and \texttt{RecSys} reap $7.22\times$, $38\times$, and $282.24\times$ improvements in energy-delay product, respectively. Note that we do not have accelerators for backpropagation of convolutional layers and therefore the modest improvement for ConvNet is due to forward propagation acceleration in the context of the entire training. In addition, GraphSage includes random walk and embedding steps that are not handled by an accelerator. RecSys on the other hand is entirely handled by accelerators and results in its impressive improvement.
These results highlight the performance benefits of accelerators for compute-bound kernels in DNN applications. 
\projname{} supports detailed accelerator performance modeling suitable for Keras TensorFlow kernels.

\section{Related Work}
\label{sec:background}
\label{sec:related-work}



Previous work on simulators, i.e. Graphite~\cite{graphite}, Sniper~\cite{sniper}, ZSim~\cite{zsim} and PriME~\cite{prime},  has focused on increasing (1)~accuracy through tuned core models and native execution and (2)~simulation scalability by executing simulations on parallel, multicore host machines. ZSim and PriME were designed to make many-core, i.e. thousands of cores, simulation practical. Furthermore, some of these works allow for flexible memory hierarchies. However, all of these prior works only simulate homogeneous core systems.

Sniper extended Graphite by providing a more detailed core model and using interval simulation. However, this results in a trade-off between accuracy and simulator performance. This level of abstraction lies in between 1-IPC models and highly detailed hardware pipelines. \projname{} also sits at this abstraction level, but makes use of LLVM IR to create a DDG for instruction scheduling in cycle-driven simulation. The use of LLVM IR also allows natural additions of compiler passes and new instructions (e.g.\ DAE in Section~\ref{sec:DAE-case-study}). 

ZSim is a many-core simulator that instruments a binary based on every basic block and memory operation. It leverages the host machine to perform functional simulation and model timing, hindering its ability to simulate different types of cores. \projname{} also performs code instrumentation and native execution for memory access behavior and control-flow resolution, but its modular, tile-based nature allows it to simulate a variety of tiles in a heterogeneous system. 

PriME was designed for many-core system simulation as well, but focuses on microarchitectural exploration, including cache hierarchies, coherence protocols, and NoCs. Though it presents a tile-based architecture like \projname{}, their tiles require homogeneity, making it an unsuitable simulator for modeling accelerator-oriented many-core systems.

\textbf{Accelerator Simulation:} Other works have focused on simulating accelerator performance. Rogers et al.~\cite{llvm_accel_gem5} devised an LLVM-based accelerator model in gem5 that leverages a data dependency graph to simplify the simulation of a many-accelerator system. \projname{} uses the same front-end, but is not limited to accelerator modeling; it supports a variety of other SoC components, including core models (e.g. in-order and out-of-order). Furthermore, \projname{} provides tile-to-tile scratchpad communication, e.g. to support data communication schemes like DAE. Because it does not rely on gem5, \projname{} allows for greater implementation flexibility and higher simulation speed. 

Gem5-Aladdin \cite{shao_micro16} is another gem5-based approach that uses Aladdin~\cite{aladdin} for fixed-function accelerator design in the context of an SoC. Gem5-Aladdin measures the impact of DMA overload in an SoC to design accelerators in a holistic manner rather than in isolation. Additionally, the work evaluates simple accelerators where normally the input and output data fit in the local memory of the accelerator. At each invocation these accelerators execute for a few thousands of cycles, which is typically less than the overhead of their invocation from a Linux device driver.

On the other hand, \projname{} can model accelerators of any complexity, e.g. accelerators for which: (1)~communication and computation are decoupled and concurrent, (2)~input and output data do not need to fit in the local memory of the accelerator, they can be of arbitrary size. For this reason, we were able to evaluate realistic accelerator workloads in terms of size. If the accelerators are invoked for small tasks, the invocation overhead dominates the execution time and the accelerator hardly achieves any speedup with respect to general purpose cores. Our measurements of accelerator execution time on FPGA included the invocation overhead. 
Furthermore, \projname{} considers heterogeneity not only in combining accelerators with a core model, but also in providing flexible core models. Its LLVM-based approach allows natural agile development of programming models, ISA extensions, and novel architectures. 

To the best of our knowledge, \projname{} is the first simulation approach for loosely-coupled heterogeneous systems, offering flexible, early-stage exploration of hardware-software co-design approaches to design new architectures.
\section{Conclusion}
This paper presents \projname{}, a lightweight, modular simulator to flexibly explore the design space of heterogeneous systems via hardware-software co-design. \projname{} (1)~is tightly integrated with the LLVM framework, providing agile programming models, enabling full-stack approaches; (2)~provides abstract tile models capturing pragmatic microarchitectural details and specialized tile-to-tile interactions; and (3)~provides support for accelerator model integration to create complex heterogeneous systems. \projname{} is a timely contribution in the New Golden Age of Computer Architecture~\cite{goldenage}, where flexible hardware-software co-design and heterogeneity are key to performance improvements.

\section*{Acknowledgment}
We would like to thank the anonymous reviewers for their helpful feedback. 
This work was supported in part by the DARPA SDH Program under agreement No. FA8650-18-2-7862. 
This research was funded in part by the U.S.\ Government. 
Prof.\ Aragón has been supported by the Spanish State Research Agency under grant TIN2016-75344-R (AEI/FEDER, EU) and by Fundación Séneca-Agencia de Ciencia y Tecnología  de la Región de Murcia, Programa Jiménez de la Espada (grant 20580/EE/18).
The views and conclusions contained herein are those of the authors and should not be interpreted as representing the official policies or endorsements, either expressed or implied, of DARPA or the U.S.\ Government.

\bibliographystyle{IEEEtran}
\bibliography{IEEEabrv,references}
\end{document}